\documentclass[11pt]{article}
\usepackage{moriond,epsfig}

\def\Journal#1#2#3#4{{#1} {\bf #2}, #3 (#4)}

\def\NIMA{{\em Nucl. Instrum. Methods} A}
\def\NPB{{\em Nucl. Phys.} B}

\def\PRL{\em Phys. Rev. Lett.}
\def\PRD{{\em Phys. Rev.} D}

\def\be{\begin{equation}}
\def\ee{\end{equation}}
\def\bea{\begin{eqnarray}}
\def\eea{\end{eqnarray}}

\begin{document}
\vspace*{4cm}
\title{RESULTS FROM HARP \\ AND THEIR IMPLICATIONS FOR NEUTRINO PHYSICS}

\author{ BORIS A. POPOV}
\address{Dzhelepov Laboratory of Nuclear Problems, JINR, 
  Joliot-Curie 6, 141980, Dubna \\ {\bf (on behalf of the HARP Collaboration)} }

\maketitle\abstracts{
\vspace*{-0.275cm}
Recent results from the HARP experiment 
on the measurements of the double-differential production cross-section
of 
pions
in proton interactions with 
beryllium, carbon and tantalum targets
are presented.
These results are relevant for a detailed understanding of neutrino
flux
in accelerator neutrino experiments 
MiniBooNE/SciBooNE, for a better prediction of atmospheric neutrino fluxes 
as well as for 
an optimization 
of a future neutrino factory design. 
}

\section{The HARP experiment}
\vspace*{-0.1cm}

The HARP experiment~\cite{ref:harp-prop,ref:harpTech} at the CERN PS
was designed to make measurements of hadron yields from a large range
of nuclear targets and for incident particle momenta from 1.5~GeV/c to 15~GeV/c.
The main motivations are the measurement of pion yields for a quantitative
design of the proton driver of a future neutrino factory, a
substantial improvement in the calculation of the atmospheric neutrino
flux
and the measurement of particle yields as input for the flux
calculation of accelerator neutrino experiments,
such as K2K~\cite{ref:k2k,ref:k2kfinal},
MiniBooNE~\cite{ref:miniboone} and SciBooNE~\cite{ref:sciboone}.

 The HARP experiment
 makes use of a large-acceptance spectrometer consisting of a
 forward and large-angle detection system.
 A detailed
 description of the experimental apparatus can be found in Ref.~\cite{ref:harpTech}.
 The forward spectrometer --- 
 based on large area drift chambers~\cite{ref:NOMAD_NIM_DC} and a dipole magnet
 complemented by a set of detectors for particle identification (PID): 
 a time-of-flight wall~\cite{barichello} (TOFW), a large Cherenkov detector (CHE) 
 and an electromagnetic calorimeter  ---
 covers polar angles up to 250~mrad which
 is well matched to the angular range of interest for the
 measurement of hadron production to calculate the properties of
 conventional neutrino beams.
 The large-angle spectrometer --- based on a Time Projection Chamber (TPC) 
 located inside a solenoidal magnet ---
 has a large acceptance in the momentum
 and angular range for the pions relevant to the production of the
 muons in a neutrino factory.
 It covers the large majority of the pions accepted in the focusing
 system of a typical design.
 The neutrino beam of a neutrino factory originates from
 the decay of muons which are in turn the decay products of pions.

\section{Results obtained with the HARP forward spectrometer}
\vspace{-0.15cm}

The first HARP physics publication~\cite{ref:alPaper} reported measurements of the
$\pi^+$ production cross-section from an aluminum target 
at 12.9~GeV/c proton momentum. 
This corresponds to the energies of the KEK PS
and the target material used by the K2K experiment.  
The results obtained in Ref.~\cite{ref:alPaper} were
subsequently applied to the final neutrino oscillation analysis 
of K2K~\cite{ref:k2kfinal}, allowing a significant reduction 
of the dominant systematic error associated with the calculation of
the so-called far-to-near ratio 
(see~\cite{ref:alPaper} and~\cite{ref:k2kfinal} for a detailed discussion) 
and thus an increased K2K sensitivity to the oscillation signal. 

A detailed description of established experimental techniques 
for the data analysis in the HARP forward spectrometer 
can be found in Ref.~\cite{ref:alPaper,ref:pidPaper}.
Our next goal is to contribute to the understanding of 
the MiniBooNE and SciBooNE neutrino fluxes. 
They are both produced by the Booster Neutrino Beam at 
Fermilab which originates from protons accelerated to 8.9~GeV/c by 
the booster before being collided against a beryllium target. 
As was the case for the K2K beam, a fundamental input for the calculation 
of the resulting $\nu_\mu$ flux is the measurement of the $\pi^+$ cross-sections 
from a thin 5\% nuclear interaction length ($\lambda_{\mathrm{I}}$)
beryllium target at 8.9~GeV/c proton momentum, which is presented here 
and in the forthcoming HARP publication~\cite{ref:bePaper}.

With respect to our first published physics paper~\cite{ref:alPaper}, 
a number of 
improvements to the analysis techniques 
and detector simulation have been made. 
The most important improvements introduced in this analysis 
compared with the one presented in Ref.~\cite{ref:alPaper} are:
\begin{itemize}
\item An increase 
of the track reconstruction efficiency which is now constant over a
much larger kinematic range
and a better momentum resolution 
coming from improvements in the tracking algorithm; 
\item Better understanding of the momentum scale and resolution of the detector, 
based on data, which was then used to tune the simulation. 
This results in smaller systematic errors associated with the unsmearing corrections 
determined from Monte Carlo; 
\item New 
particle identification
hit selection algorithms both in 
the 
TOFW
and in the 
CHE
resulting 
in much reduced background and negligible efficiency losses; 
\item Significant increases in Monte Carlo production  
have also reduced uncertainties from Monte Carlo statistics 
and allowed studies which have reduced certain systematics.
\end{itemize}
It is important to point out that an analysis incorporating 
these improvements yields results for the aluminum data fully consistent 
with those published in Ref.~\cite{ref:alPaper}.

The absolutely normalized double-differential cross-section for the process 
$p + \mbox{Be} \rightarrow \pi^+ + X$
can be expressed in bins of pion kinematic variables in the
laboratory frame, ($p_{\pi},\theta_{\pi}$), as 
\begin{equation}
  \frac{d^2\sigma^{\pi^{+}}}{dpd\Omega}(p_{\pi},\theta_{\pi}) = \frac{A}{N_{\mbox{A}}\rho t}\cdot\frac{1}{\Delta p\Delta\Omega}\cdot\frac{1}{N_{\mbox{pot}}}
    \cdot N^{\pi^{+}}(p_{\pi},\theta_{\pi}) \ ,
    \label{eq:truexsec}
\end{equation}
where:
\begin{itemize}
\item  $\frac{d^2\sigma^{\pi^{+}}}{dpd\Omega}$ is the cross-section in 
\ensuremath{\mbox{cm}^2/(\mbox{GeV/c})/\mbox{srad}} 
for each ($p_{\pi},\theta_{\pi}$) bin covered in the analysis.
\item $\frac{A}{N_{\mbox{\tiny A}}\rho}$ is the reciprocal of the number density of
  target nuclei for 
Be
($1.2349 \cdot 10^{23}$ per cm$^3$).
\item $t$ is the thickness of the beryllium target along the beam
  direction. 
       The thickness is measured to be 2.046~cm with a maximum
       variation of 0.002~cm.
\item $\Delta p$ and $\Delta \Omega$ are the bin sizes in momentum and solid angle, respectively.\footnote{$\Delta p = p_{max}-p_{min};
\ \Delta \Omega = 2\pi(cos(\theta_{min}) - cos(\theta_{max}))$} 
\item $N_{\mbox{pot}}$ is the number of protons on target after event
  selection cuts.
\item $N^{\pi^{+}}(p_{\pi},\theta_{\pi})$ is the yield of positive pions in bins
  of true momentum and angle in the laboratory frame.% ($p_{\pi},\theta_{\pi}$). 
\end{itemize}
Eq.~\ref{eq:truexsec} can be generalized to give the inclusive
cross-section for a particle of type $\alpha$      
\begin{equation}
  \frac{d^2\sigma^{\alpha}}{dpd\Omega}(p,\theta) = \frac{A}{N_{\mbox{A}}\rho t}\cdot\frac{1}{\Delta p\Delta\Omega}\cdot\frac{1}{N_{\mbox{pot}}}
  \cdot M^{-1}_{p\theta\alpha p^{'}\theta^{'}\alpha^{'}}\cdot N^{\alpha^{'}}(p^{'},\theta^{'}) \ ,
  \label{eq:recxsec}
\end{equation}
where reconstructed quantities are marked with a prime and 
$M^{-1}_{p\theta\alpha p^{'}\theta^{'}\alpha^{'}}$ is the
inverse of a matrix which fully describes the migrations between bins
of true and 
reconstructed quantities, namely: lab frame momentum, p, lab frame
angle, $\theta$, and particle type, $\alpha$.

There is a background associated with beam protons interacting in materials 
other than the nuclear target (parts of the detector, air, etc.).  
These events 
are
subtracted by using data collected without the nuclear target 
in place where one has been careful to normalize the sets to the same 
number of protons on target. This procedure is referred to as 
the `empty target subtraction':  
\begin{equation}
  N^{\alpha^{'}}(p^{'},\theta^{'}) \rightarrow [N^{\alpha^{'}}_{\mbox{target}}(p^{'},\theta^{'}) - N^{\alpha^{'}}_{\mbox{empty}}(p^{'},\theta^{'})] \ .
\end{equation}

The event selection is performed in the following way:
a good event is required to have a single, well reconstructed and
identified beam particle 
impinging on the nuclear target.  
A downstream trigger in the forward trigger plane (FTP) 
is also required to record the event, necessitating an additional set of unbiased,
pre-scaled triggers for absolute normalization of the  
cross-section.  These pre-scale triggers (1/64 for the 8.9~GeV/c Be
data set) are subject to exactly the same selection 
criteria for a `good' beam particle as the event triggers allowing the
efficiencies of the selection to cancel, thus adding no additional
systematic uncertainty to the absolute normalization of the result. 
Secondary track selection criteria have been optimized to ensure the
quality of the  momentum reconstruction as well as a clean time-of-flight 
measurement while maintaining high reconstruction and particle identification 
efficiencies.
The results of the event and track selection in the beryllium thin target data
set are shown in Table~\ref{tb:be5events}.

\begin{table} 
\caption{Total number of events in the 8.9~GeV/c beryllium
  5\%~$\lambda_{\mathrm{I}}$ target 
  and empty target data sets, and the number of protons on target as
  calculated from the prescaled trigger count.} 
\label{tb:be5events}
\begin{center}
\begin{tabular}{| l | c | c |} \hline
\bf{Data Set} & \bf{8.9 GeV/c Be 5\% $\lambda_{\mathrm{I}}$} & \bf{8.9 GeV/c Empty Target}\\ \hline
    protons on target                & 13,074,880 & 1,990,400 \\
    total events processed           & 4,682,911 & 413,095\\ 
    events with accepted beam proton & 2,277,657 & 200,310 \\ 
    beam proton events with FTP trigger & 1,518,683 & 91,690 \\
    total good tracks in fiducial volume & 95,897  & 3,110 \\ \hline 
\end{tabular}
\end{center}
\end{table}

\begin{figure}
  \begin{center}
    \includegraphics[width=12cm,height=8cm]{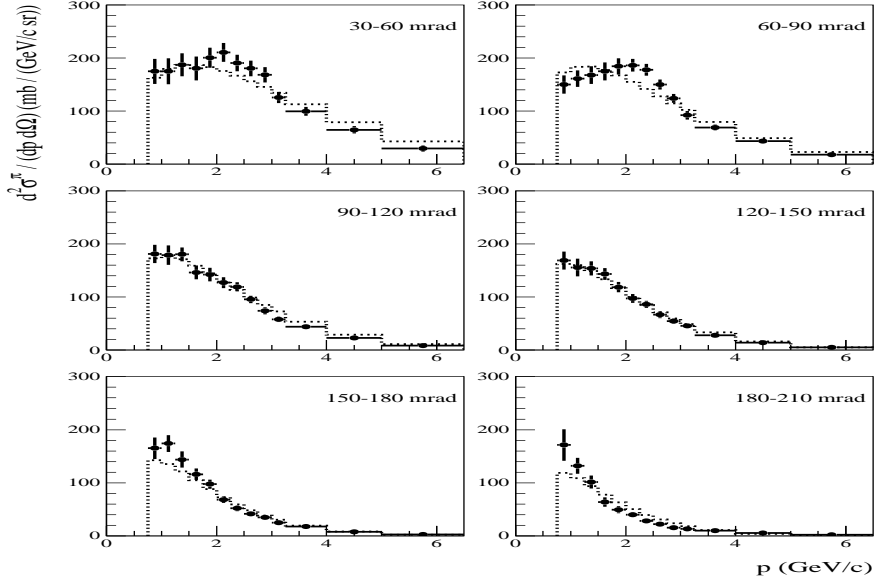}
    \caption{\label{fig:xsecBins}
      HARP measurements of the double-differential production
      cross-section of positive pions, $d^2\sigma^{\pi^{+}}/dpd\Omega$,
      from 8.9~GeV/c protons on 5\%~$\lambda_{\mathrm{I}}$ beryllium target 
      as a function 
      of pion momentum, p, in bins of pion angle, $\theta$, in the 
      laboratory frame.  The error bars shown include statistical
      errors and all (diagonal) systematic errors. The dotted histograms show 
      the Sanford-Wang parametrization that best fits the HARP data.
    }  
  \end{center}
\end{figure}

The double-differential inelastic cross-section for the
production of positive pions from collisions of 8.9~GeV/c protons 
with beryllium have been measured in the kinematic range from 
$0.75 \ \mbox{GeV/c} \leq p_{\pi} \leq 6.5$~GeV/c and 
$0.030 \ \mbox{rad} \leq \theta_{\pi} \leq 0.210$~rad,
subdivided into 13 momentum and 6 angular bins.  
Systematic errors have been estimated.  
A full $(13 \times 6)^2 = 6048$ element covariance matrix has been generated
to describe the correlation among bins. The data are presented graphically as 
a function of momentum in 30 mrad bins in Fig.~\ref{fig:xsecBins}.  
To characterize the uncertainties on this measurement we show the diagonal
elements of the covariance matrix plotted on the data points in Fig.~\ref{fig:xsecBins}. 
A typical total uncertainty of 9.8\% on the double-differential cross-section values 
and a 4.9\% uncertainty on the total integrated cross-section are obtained.

Sanford and Wang~\cite{ref:SW} have developed an empirical parametrization for
describing the production cross-sections of mesons in proton-nucleus interactions.
This parametrization has the functional form: 
\begin{equation}
\label{eq:swformula}
\frac{d^2\sigma (\hbox{p+A}\rightarrow \pi^++X)}{dpd\Omega}(p,\theta) =
 \exp [ c_{1}-c_{3}\frac{p^{c_{4}}}{p_{\hbox{\footnotesize beam}}^{c_{5}}}-c_{6}
 \theta (p-c_{7} p_{\hbox{\footnotesize {beam}}} \cos^{c_{8}}\theta ) ] p^{c_{2}}
 (1-\frac{p}{p_{\hbox{\footnotesize beam}}}) \ ,
\end{equation} 
where $X$ denotes any system of other particles in the final state,
$p_{\hbox{\footnotesize {beam}}}$ is
the proton beam momentum in GeV/c, $p$ and $\theta$ are the $\pi^+$
momentum and angle in units of GeV/c and radians, respectively, 
$d^2\sigma/(dpd\Omega)$ is expressed in units of mb/(GeV/c\ sr),
$d\Omega\equiv 2\pi\ d(\cos\theta )$, 
and the parameters $c_1,\ldots ,c_8$
are obtained from fits to meson production data.

The $\pi^+$ production data reported here have been fitted 
to this empirical formula (Eq.~\ref{eq:swformula}).
In the $\chi^2$ minimization, the full error matrix was used. 
The best-fit values of the Sanford-Wang parameters are reported 
in Table~\ref{tab:swpar_values_errors}, together with their errors.

\begin{table}%[tb]
\caption{\label{tab:swpar_values_errors}
Sanford-Wang parameters and errors obtained by fitting the
  dataset. The errors refer to the 68.27\% confidence
 level for seven parameters ($\Delta\chi^2=8.18$).
}
{
\tiny
\begin{center}
\begin{tabular}{| c | c | c | c | c | c | c | c |} \hline
{\bf Parameter} & $c_1$ & $c_2$ & $c_3$ & $c_4=c_5$ & $c_6$ & $c_7$ & $c_8$ \\ \hline
{\bf Value} & $(82.2\pm 19.8)$ & $(6.47\pm 1.62)$ & $(90.6\pm 20.3)$ &
 $(7.44\pm 2.30)\times 10^{-2}$ & $(5.09\pm 0.49)$ & $(0.187\pm 0.053)$ &
 $(42.8\pm 13.6)$ \\ \hline
\end{tabular}
\end{center}
}
\end{table}

The MiniBooNE neutrino beam is produced from the decay of $\pi$ and K mesons 
which are produced in collisions of 8.9~GeV/c protons from the Fermilab Booster 
on a 71 cm beryllium target. The neutrino flux prediction is generated 
using a Monte Carlo simulation.
In this simulation the primary meson production rates
are taken from a fit of existing data with a Sanford-Wang empirical 
parametrization
in the relevant region. 
The results presented here, being for protons at exactly the booster beam energy, 
are then a critical addition to these global 
fits.
The kinematic region of the measurements presented
here contains 80.8\% of the pions contributing 
to the neutrino flux in the MiniBooNE detector. 

\begin{figure}
  \begin{center}
    \includegraphics[width=16cm,height=7.5cm]{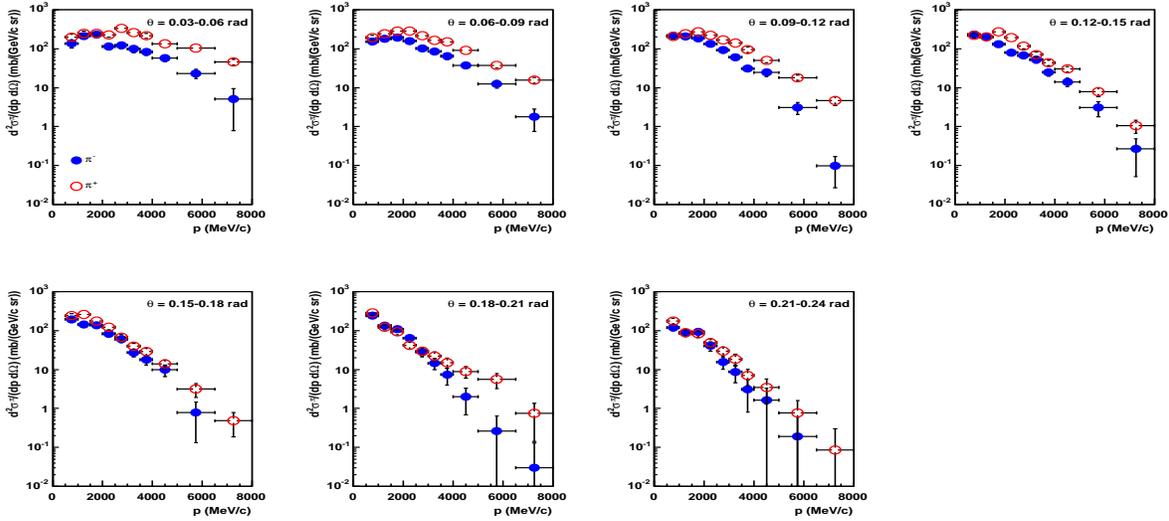}
    \caption{\label{fig:fw_carbon}
      Measurements of the double-differential production
      cross-sections of 
      $\pi^+$ (open circles) and $\pi^-$ (closed circles)
      from 12~GeV/c protons on 5\%~$\lambda_{\mathrm{I}}$ carbon target as a function 
      of pion momentum, p, in bins of pion angle, $\theta$, in the 
      laboratory frame.  The error bars shown include statistical
      errors and all (diagonal) systematic errors. 
    }  
  \end{center}
\end{figure}

A similar analysis has been performed using the HARP forward spectrometer 
for the measurement 
of the double-differential production cross-section of $\pi^\pm$ 
in the collision of 12~GeV/c protons with a thin 5\%~$\lambda_{\mathrm{I}}$ 
carbon target.
The results are shown in Fig.~\ref{fig:fw_carbon}.
These measurements are important for a precise calculation of the atmospheric neutrino
flux and for a prediction of the development of extended air showers.

\vspace*{-0.35cm}
\section{Results obtained with the HARP large-angle spectrometer}
\vspace*{-0.15cm}

\begin{figure}[tbp]
\begin{center}
\epsfig{figure=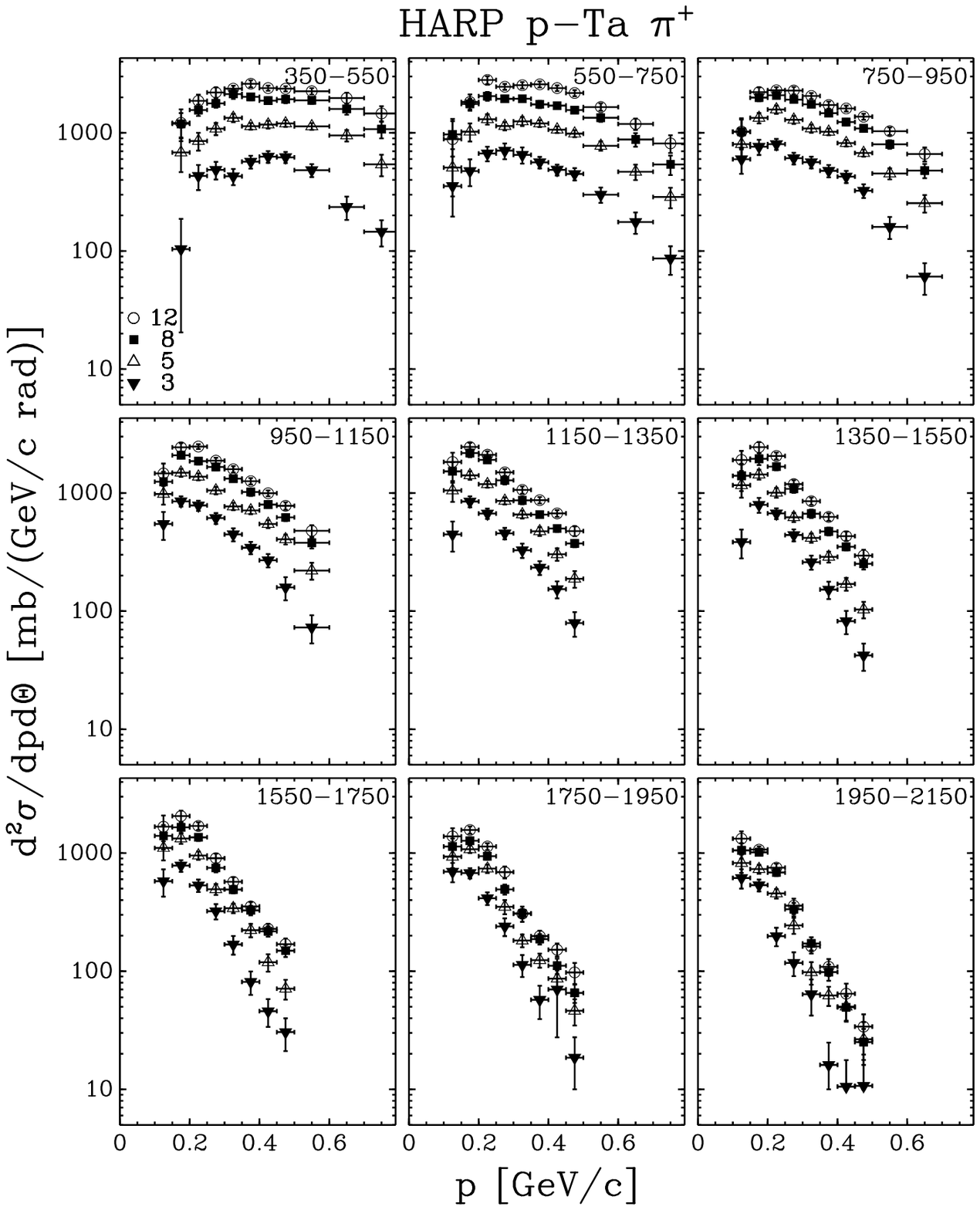,height=8cm,width=0.49\textwidth} 
\epsfig{figure=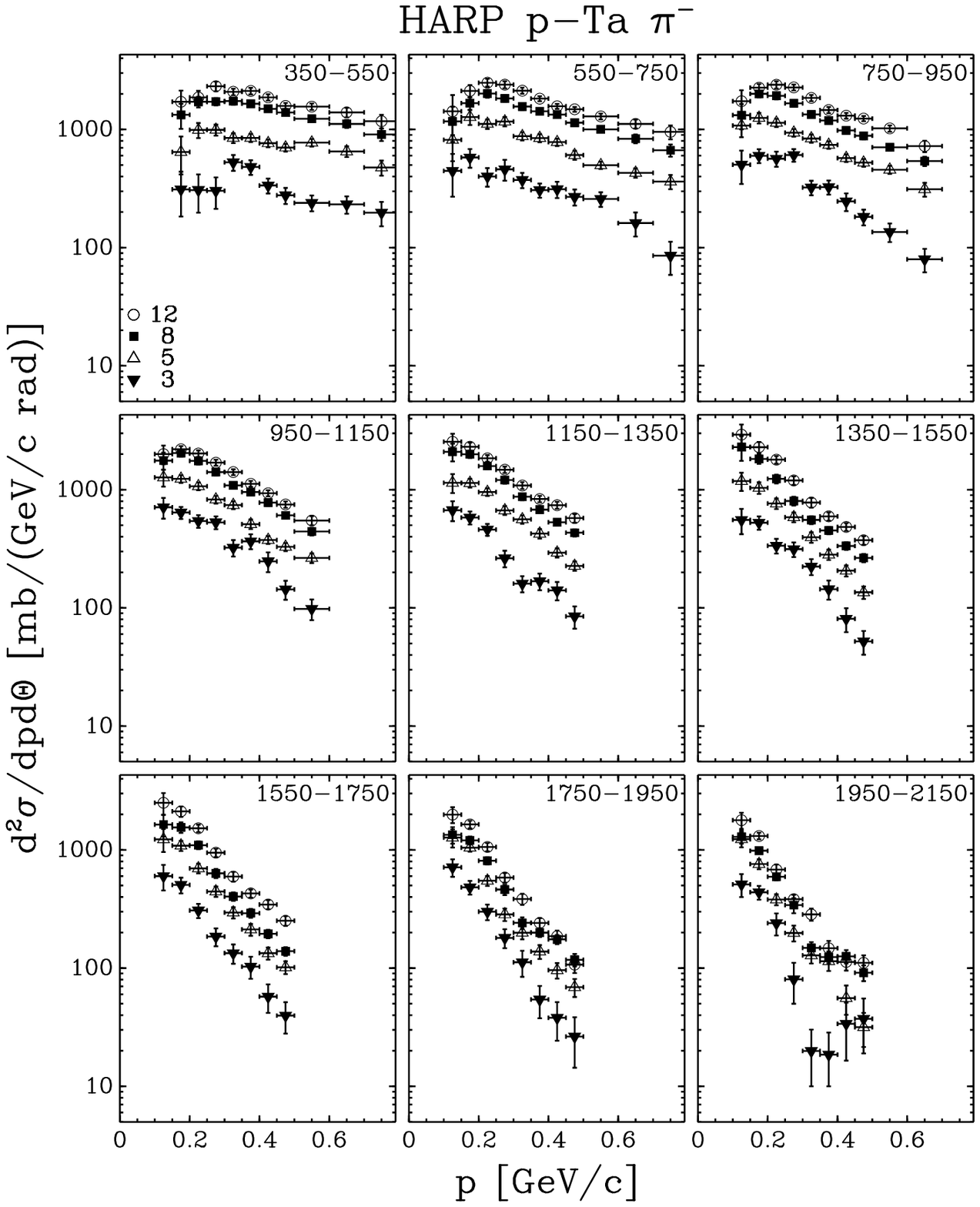,height=8cm,width=0.49\textwidth} 
\caption{
Double-differential cross-sections for $\pi^+$ (left) and $\pi^-$ (right) 
production in p--Ta interactions as a function of momentum displayed in different
angular bins (shown in mrad in the panels).
The results are given for all incident beam momenta (filled triangles:
3~GeV/c; open triangles: 5~GeV/c; filled rectangles: 8~GeV/c; open
circles: 12~GeV/c). 
The error bars take into account the correlations of the systematic
 uncertainties. 
}
\label{fig:la_tantalum}
\end{center}
\end{figure}

  First results on the measurements of 
  the double-differential cross-section for the production
  of charged pions in proton--tantalum collisions emitted at large
  angles from the incoming beam direction have been obtained 
  recently~\cite{ref:TaPaper}. 
  The pions were produced by proton beams in a momentum range from
  3~GeV/c to  12~GeV/c hitting a tantalum target with a thickness of
  5\%~$\lambda_{\mathrm{I}}$.  
  The angular and momentum range covered by the experiment 
  ($100~\mbox{MeV/c} \le p < 800~\mbox{MeV/c}$ and 
  $0.35~\mbox{rad} \le \theta <2.15~\mbox{rad}$)
  is of particular importance for the design of a neutrino factory.
  Track recognition, momentum determination and particle
  identification were all performed based on the measurements made with
  the TPC. 
  Results for the double-differential cross-sections 
  $
  {{\mathrm{d}^2 \sigma}}/{{\mathrm{d}p\mathrm{d}\theta }}
  $
  at four incident proton beam momenta (3~GeV/c, 5~GeV/c, 8~GeV/c 
  and 12~GeV/c) are shown in Fig.~\ref{fig:la_tantalum}.  

Similar analyses are being performed for the Be, C, Cu, Sn and Pb targets using 
the same detector, which will allow a study of A-dependence of the pion yields 
with a reduced systematic uncertainty to be performed.

\vspace*{-0.35cm}
\section{Conclusions}\label{sec:conclusions}
\vspace*{-0.25cm}

Measurements of the double-differential production cross-section
of positive pions in the collision of 8.9~GeV/c protons 
with a beryllium target have been presented.  The data have been
reported in bins of pion momentum and angle in the kinematic
range from $0.75 \ \mbox{GeV/c} \leq p_{\pi} \leq 6.5$~GeV/c and 
$0.030 \ \mbox{rad} \leq \theta_{\pi} \leq 0.210 \ \mbox{rad}$.   
A systematic error analysis has been performed yielding 
an average point-to-point error of 9.8\% (statistical + systematic) and
an overall normalization error of 2\%.  
The data have been fitted to the empirical parameterization
of Sanford and Wang and the resulting parameters provided.    
These production data have direct relevance for the prediction of a 
$\nu_\mu$ flux for MiniBooNE
and SciBooNE experiments.

Preliminary results for the measurement 
of the double-differential production cross-section of $\pi^\pm$ 
in the collision of 12~GeV/c protons with a carbon target have been presented.

First results on the production of pions at
large angles with respect to the beam direction for protons of
3~GeV/c, 5~GeV/c, 8~GeV/c and 12~GeV/c impinging on a thin
tantalum target have been described. 
These data can be used to make predictions for the fluxes of pions to
enable an optimized design of a future neutrino factory.

\vspace*{-0.5cm}
\section*{Acknowledgments}
\vspace*{-0.3cm}
It is a pleasure to thank the organizers for the financial support 
which allowed me to 
participate in the conference and to present 
these results on behalf of the HARP Collaboration.

\end{document}